# NON-FUNGIBLE TOKENS IN BUSINESS AND MANAGEMENT – A REVIEW

Najam A. Anjum*, and Mubashir Husain Rehmani**

*Lancaster University, Leipzig, Germany, Email: n.anjum1@lancaster.ac.uk
** Munster Technological University (MTU), Cork, Ireland, Email: mshrehmani@gmail.com

## ABSTRACT

Non-Fungible Tokens (NFTs) are a new development in blockchain technology. News around NFTs is surrounded by skepticism because unrealistically high prices are being paid online for these NFTs which are in the form of apparently simple digital arts and photographs. It is not clear if this is a trend, a hype, a bubble, or a legitimate novel way of holding and trading value. A literature review of peer-reviewed scholarly studies, performed in the context of business and management, is presented here. Moreover, we also discuss open issues, and challenges, and present future research directions. Analysis of these studies reveal that schools of thoughts are divided on the validity of this form of digital tokens. On one hand, there is a lot of criticism but on the other hand, we can find novel business models and applications of NFTs especially the feature of smart contracts. It can, therefore, be concluded that NFTs, even if not in their current form, are here to stay and may promise new ways of protecting digital assets in an immutable and easily traceable form.

# I. INTRODUCTION

Non-Fungible Tokens or NFTs are fast gaining popularity. Currently, the most popular use of NFTs, within the domain of business and management, can be seen in the field of digital art but there have also been attempts at using this tokenization for securing real estate, music (Chalmers et al, 2022), and other forms of intellectual property (Fairfiled, 2021). Although, sales of digital art amounting to as big as $69 million can be seen (Dunne, 2022), the talk around the legitimacy of NFTs is facing some skepticism because of the astronomical rise in prices of these digital assets within the time span of a few months and some evidence of wash-trading (Chalmers et al, 2022). There can, however, be seen some work developing around the use and application of NFTs in other fields as well (Regner et al, 2019; Popescu, 2021; Dowling, 2021a; 2021b).

Because of being a relatively recent phenomenon, the scholarly work around NFTs is still immature and is at a stage where researchers are grappling with the issues of correct understanding and future scope of this technology especially in the field of business and management. In this context, Andrea et al (2022) discussed the impact of NFTs on consumers and marketing strategies. Authors first provided a detailed discussion on marketing literature on NFTs. Then authors focused on how materialism, status consumption orientation, and innovativeness are related with the buying of NFTs in artworks, music compilations, and fashion products. Another interesting review article is on the pricing aspect of NFTs (Bao et al, 2022), however, authors only covered 13 articles published in economics and finance journals in their review.

Complementary to these works, in this study therefore, the authors contribute to the growing set of research literature that attempts at helping the research community in making sense of this concept. Hence, we ask the following very basic questions:

1. Are NFTs here to stay or are they just a fad and a bubble soon to be burst.
2. What other applications of NFTs can there be if this tokenization method is considered legitimate?

The study presented here reviews the scholarly work being done on NFTs and to answer these questions. This is such a new field of research that not many peer-reviewed studies can be found. The aim of this study, therefore, is to establish an understanding of this field and identify possible research directions that it may take in the times to come. In the following sections, NFTs, their characteristics, and market trends have been discussed. We then discuss in detail the use of NFTs in business and management, finance, and social sciences. Finally, we highlight issues, challenges, and future research direction along with providing conclusions obtained through the findings of this review.

## II. WHAT ARE NON-FUNGIBLE TOKENS?

NFTs are tradeable rights to digital assets. These rights are created through the establishment of a smart contract on the Ethereum blockchain (Dowling, 2021 a). Non-Fungible Tokens came into inception through the ERC-721 standard in 2017 (Entriken et al, 2018) which prescribes an improved way of tokenization of individual assets which was not possible with fungible tokens because of their inability to represent uniqueness (Regner et al, 2019). Being non-fungible means a particular NFT has a unique value. This means these digital assets cannot be traded one-on-one like other fungible cryptocurrencies e.g., Bitcoin, Ether, etc (Popescu, 2021). Figure 1 shows types, characteristics, and usage of NFTs.

**Figure 1: Types, Characteristics, and Usage of NFTs**

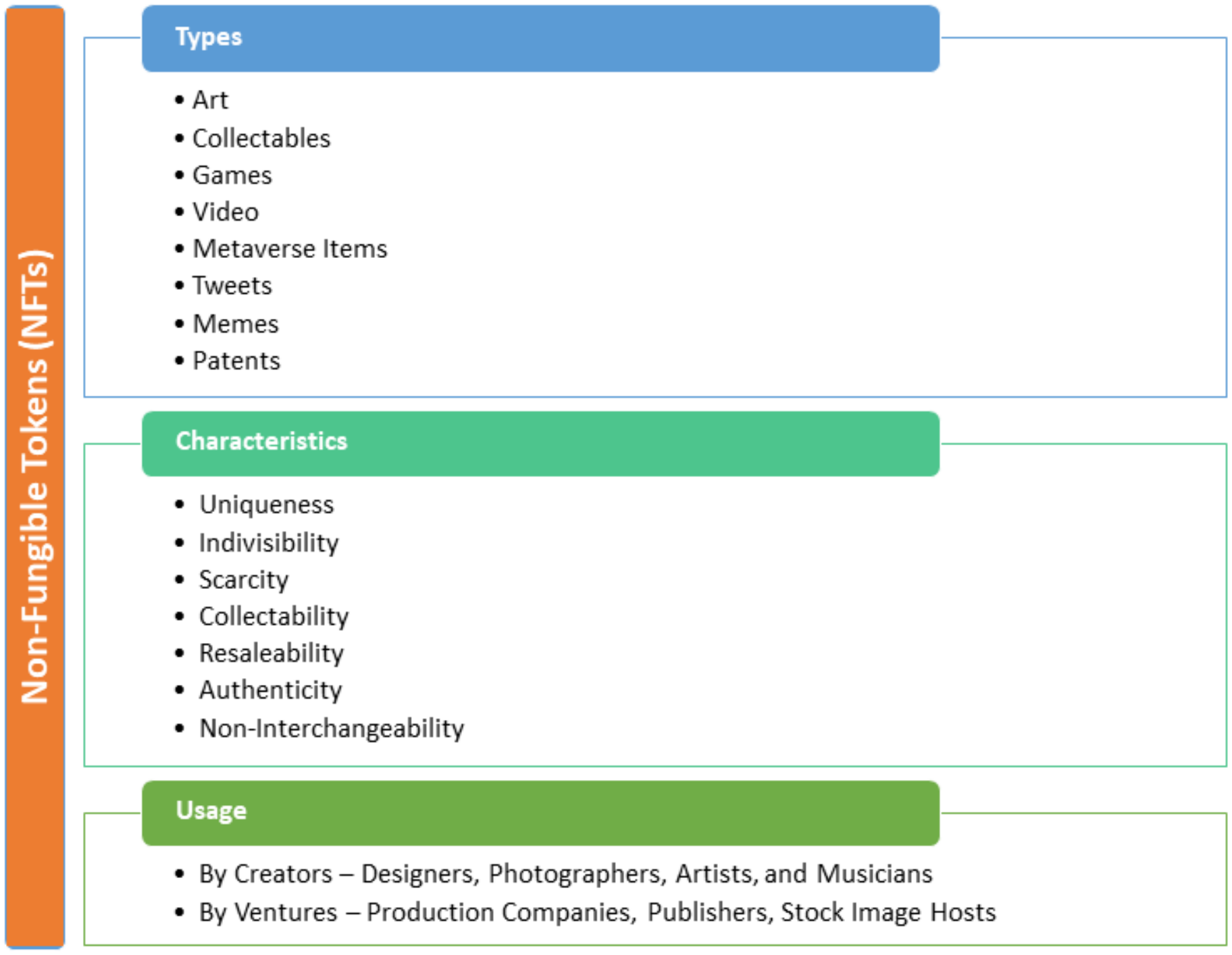

### ***Possible applications***

NFTs, in a way, provide a Proof-of-Asset, its indivisible and hence cannot be divided into smaller components, its accurately verifiable since it is connected to a blockchain, and its record is immutable for the same reason (Popescu, 2021). It also improves the tokenization of individual unique assets which is not possible in fungible tokens (Regner et al, 2019). NFTs can be programmed to pay royalties, they can provide the records of ownership for in-game items, they can be used to prevent digital content piracy over the Internet and can also be used as collateral to borrow Stablecoins like DAI (Popescu, 2021).

### *Market Trends in sale of digital art*

As understood, NFTs could be useful in connecting the digital tokenization with the market of collectibles, which according to an estimate, have a global market size of USD 200 billion (Fenech, 2018). The first notable NFT based application that gained prominence was an online game called 'Crypto Kitties' which became so popular that at one point it took more than 70% of network capacity of Ethereum blockchain and the NFTs representing ownership of these cat characters got sold for as high as USD 100,000 (Tepper, 2017).

Some of the high-valued NFTs are listed in Table 1. As can be seen, 'Everydays: The first 5000 days' was sold for $69.3m. This purchase took place in March 2021 and the buyer, Vignesh Sundaresan, nicknamed MetaKovan, a Singaporean citizen and an active entrepreneur, seems convinced that digital artwork NFTs could be a useful medium to hold and trade value especially in the times of Metaverse (Frank, 2021).

**Table 1: Some of the most expensive NFTs ever sold**

| S. No | NFT Artwork | Selling price (million USD) |
|---|---|---|
| 1 | Everydays: The first 5000 days | 69.3 |
| 2 | CryptoPunk #7523 | 11.8 |
| 3 | CryptoPunk #7804 | 7.56 |
| 4 | CryptoPunk #3100 | 7.51 |
| 5 | Crossroads | 6.66 |
| 6 | Ocean Front | 6 |
| 7 | CryptoPunk #5217 | 5.59 |
| 8 | World Wide Web Source Code | 5.4 |
| 9 | Save Thousands of Lives | 5.23 |
| 10 | Stay Free | 5.27 |

Data Source: Dunne (2022)

## III. NFTs IN BUSINESS, MANAGEMENT, AND ACCOUNTING

In this section, we discuss the use of NFTs in Business and Management. Table 2, 3, and 4 list the papers that deliberate the applications of NFTs in Business and Management. Scopus database was used to find these papers. Since this is a new and emerging field, very few studies from reputable sources could be found.

One of the latest works, when searching for NFTs, comes out to be that of Whitaker (2022). Here, the author does narrate a case created out of interviews but upon detailed review this publication doesn't appear to fall in the category of research focusing on NFTs. The work of Bouraga (2021), Ismail et al (2021), and Casale-Brunet et al (2021) attempts at exploring the integration and resulting effects of NFT in the existing socio-economic fabric. Another growing set of research literature on NFTs deals with its applications in different fields. This is apparent from the work of Ismail et al (2021) and Mofokeng and Matima (2018) who analyze the applicability of NFTs in the fields of supply chain management and wildlife conservation respectively. The work of Mofokeng and Matima (2018) also discusses the use of NFTs in Metaverse. The echo of this application in the virtual worlds can also be heard in other fields of management science as will be discussed later. Regner et al (2019) demonstrate the use of NFTs in event ticketing. They design, prototype, and implement a decentralized NFT-based event-ticketing system. They found that this system could be a good replacement of a conventional event-ticketing system because it may help in preventing fraud, better control on secondary market transactions, and validation of ownership.

Another important point that can be noticed in the literature is the positive connotation around the use of Smart Contracts that binds digital assets with Ethereum blockchain. It is this feature of Ethereum that makes it useful for applications in supply chain management as highlighted by Ismail et al (2021). Another application of NFTs that is also apparent in the other fields of management sciences is its use in Intellectual Property protection. A glimpse of this can be seen in the work of Whitaker (2022) and Chalmers et al (2022) and more is visible in the search results of 'Economics, Econometrics, and Finance' and 'Social Sciences'. To cap the review of research in business and management, the work of Chalmers et al (2022) must be discussed which specifically focusses on the entrepreneurial potential of NFTs. The authors do show some excitement for this new and emerging technology, especially its application in protecting digital rights for artwork, music, and properties, but are also cautious of its widespread use as they highlight the presence of wash trade and an artificial hype that, as always, may lead to the bursting of this bubble and thus making a big correction in the pricing and value of NFTs. History of frauds related to Initial Coin Offering (ICO) in the cryptocurrency space also make the trade around NFTs a little questionable. Any investment in this area should, therefore, be governed by calculated risks.

A framework about NFT enabled entrepreneurship has been proposed by Chandra, Y. (2022). In the proposed framework, technology enablers as well as mechanisms were discussed in detail. From the marketing perspective, Chohan, R. (2021) presented a case study and identified enormous potential applications for marketers in NFT domain.

## IV. NFTs IN ECONOMICS, ECONOMETRICS, AND FINANCE

Search results of the use of NFTs in Economics, Econometrics, and Finance, are presented in Table 3. It can be seen that, apart from the study of van Haaften-Schick and Whitaker (2022), all other studies are finance related.

**Table 2: Search results at Scopus for 'Business and Management'**

| S. No. | Authors | Scope | Methodology | Relevant Findings and Conclusions |
|---|---|---|---|---|
| 1 | Chalmers et al (2022) | Identification of potential opportunities in NFT for creative entrepreneurs | Literature review | NFTs should be approached with caution |
| 2 | Whitaker (2022) | Proposing a conceptual framework of ungenerous and generous contracts to avoid draconian practices | Case Study | Nothing relevant to NFTs |
| 3 | Bouraga (2021) | Proposing a correlation analysis between various NFTs' characteristics and the popularity of the NFTs | Heat map analysis of transfer record data collected for top 500 NFTs up on sale online | A strong positive correlation between Total Transfers and Holders and between Total Transfers and Supply of NFTs |
| 4 | Ismail et al (2021) | Demonstrating the usefulness of smart contracts for supply chain access control and for virtual assets in a supply chain | Model building and testing through prototyping | Smart contract driven tokenization show promise in intelligently controlling access rights of different roles in a supply chain |
| 5 | Casale-Brunet et al (2021) | Systematic analysis of transactions in an NFT ecosystem based on Ethereum blockchain | Analysis of cluster and line graphs created out of transaction data extracted from the blockchains | Structure of interaction in an NFT networks is similar to the structure of interactions in a social network |
| 6 | Mofokeng and Matima (2018) | Proposing the use of NFTs as tokens for unique and rare crypto-wildlife assets the trading of which can be used for wildlife conservation | Background review | Use of NFTs as virtual unique collectable crypto-wildlife assets show promise especially as tradeable tokens in metaverse |
| 7 | Chandra Y. (2022) | Proposed NFT enabled entrepreneurship (NFTE) | Proposed framework | NFT enabled entrepreneurship can open new opportunities |
| 8 | Chohan et al (2021) | Discussed how NFTs can be used by marketers | Case Study | Identified that NFTs has enormous potential in marketing |

**Table 3: Search results at Scopus for 'Economics Econometrics and Finance'**

| S. No. | Authors | Scope | Methodology | Relevant Findings and Conclusions |
|---|---|---|---|---|
| 1 | Dowling (2021a) | Finding evidence to see if NFT pricing is correlated with cryptocurrency pricing | Analysis of descriptive statistic graphs made out of pricing data obtained from online sources | NFT pricing is quite distinct in terms of volatility transmission when compared with cryptocurrency pricing and there is little spillover between NFT markets |
| 2 | Dowling (2021b) | Exploring the presence of pricing behavior in NFC markets | Analysis of descriptive statistics graphs made out of pricing data obtained from online sources | The analysis reveals an inefficiency in pricing but a sharp rise in value of NFTs |
| 3 | Aharon and Demir (2021) | Analysis of the connectedness between returns of traditional financial assets and that of NFTs | Static and dynamic analysis using TVP-VAR methodology | Static analysis reveals that NFTs do not get affected by the shocks from other traditional assets classes and even from Ethereum. Dynamic analysis suggests that in stable market conditions, NFTs act as transmitters of systemic risk while in more unstable market condition, such as Covid, NFTs act as absorbers of risk spillovers. |
| 4 | van Haaften-Schick and Whitaker (2022) | Discussion on the different forms of use of 'Smart Contracts' on Ethereum Blockchain for digital artists | Study of contemporary data on the online NFT marketplace 'Super Rare' | Smart Contracts have the potential to be a disruptive intervention in the world of art trade and IP rights |
| 5 | Umar et al (2022) | Analysis of connectedness between NFTs and other traditional asset classes | Use of squared wavelet coherence (SWC) technique on data obtained from major indexes | Return coherence depends upon the time period of investment. In this case a threshold of two weeks was observed |

**Table 4: Search results at Scopus for 'Social Sciences'**

| S. No. | Authors | Scope | Methodology | Relevant Findings and Conclusions |
|---|---|---|---|---|
| 1 | Truby et al (2022) | Environmental damage caused by cryptocurrencies and blockchain based tokens due to high energy consumption | Study of energy consumption trends of different blockchains across countries | Proof of work method should be phased out to make the processes of blockchain more sustainable |
| 2 | Zaucha and Agur (2022) | Attitudes and behaviors apparent in the people involved in trading of commodified NBA fan experience | Thematic analysis of conversation obtained from an online chatting platform called Cord | The main social force behind the hype of NFTs is a social pressure that pushes the people to invest and be a part of the game |
| 3 | Fernandez (2021) | Exploring the applications of NFT in libraries | Opinion. Research Note. | Apparently, there are differences in the nature and functionalities of libraries and NFT, the two do have a potential for interaction in the future |
| 4 | Okonkwo (2021) | Possible applications of NFTs in IP and its commercialization | Opinion, literature review | Discussed relationship and applications of NFTs in IP |
| 5 | Heim and Hopper (2022) | Exploring the applications of blockchain in circular fashion in textile | Thematic analysis of responses to semi structured interviews | The IT systems around textile industry are not ready for blockchain adoption but this technology does show promise |
| 6 | Mackenzie and Berzina (2021) | Social and criminal lives of NFTs. Are they preventing crime or causing it? | Opinion | NFTs are a hype and their status as tokens with a potential of preventing online fraud is wrong |
| 7 | Kshetri, N. (2022) | Explored frauds, scams, and deceits in the NFT market | Opinion | Emphasized the need of clear regulatory framework to reduce scams and fraud in the NFT market |

In one of his studies, Dowling (2021 b) studied the secondary market NFT trades between March 2019 and March 2021 that took place against virtual land in the largest blockchain based virtual world called Decentraland. Authors considered LAND prices in Decentraland and evaluated Automatic variance ration (AVR), Automatic portmanteau (AP), Consistent Test (DLT), and Corrected Empirical Hurst Exponent scores. This work is important because of its investigation into the application of NFTs in virtual worlds. Aharon and Demir (2021) and Umar et al (2022) do some similar work but they check the connectedness of NFTs with common asset classes and they discover mixed results. When it comes to application of NFTs, the work of van Haaften-Schick and Whitaker (2022) is the only one that contributes to this direction. In line with the study of Ismail et al (2021), van Haaften-Schick and Whitaker (2022) also highlight the usefulness of Smart Contracts and how they make NFTs applicable for digital rights for art, music, and other forms of digital properties.

## V. NFTs IN SOCIAL SCIENCES

The work of Fernandez (2021) and Heim and Hopper (2022) qualify to be application oriented as the former explores NFTs for library applications while the latter looks into the use of this technology in IT service, in circular fashion, and in textile industry. An unmistakable, and quite intriguing, aspect of studies in social sciences is their criticism for NFTs. The connotations apparent in these papers are clearly negative. Studies by Zaucha and Agur (2022) and Mackenzie and Berzina (2021) are especially relevant here. They present a sharp critique on NFTs and they conclude that NFTs do not have any intrinsic value, it's a hype, and the only reason people are highly interested in them is social pressure that pushes them to be a part of this trend. Even the work of Fernandez (2021) and Heim and Hopper (2022), which is more application oriented, seems to present its argument in a cautious tone with suspicion and doubt evident between the lines. One similar dimension of application of NFTs in intellectual property can also be seen here in the form of the work of Okonkwo (2021).

## VI. ISSUES, CHALLENGES AND FUTURE RESEACH DIRECTION

In this section, we discuss issues, challenges, and future research directions related with NFTs in the context of business, and management, accounting, economics, finance, and social sciences.

### Coupling of Physical Assets and Their Digital Representation

Buying and selling virtual piece of land is common these days for its use in Metaverse such as NextEarth, Earth2, Decentraland, and Sandbox. The NFT of this virtual piece of land mapped onto the earth can be traded and a complete record of ownership can be traced using NFTs, but currently this virtual piece of land is not coupled with the actual physical earth. In order to completely envisage the implementation of Metaverse using Augmented Reality (AR) and Virtual Reality (VR), one can think of Avatars interacting with each other, attending meetings, visiting conferences, and owning their own piece of virtual land on Metaverse, which should somehow couple with the actual physical earth. The question is how to couple physical asset (digital representation of earth land) with the actual physical land? Who manages it? Note that it will be against the main theme of decentralization which blockchain and NFT will bring that a central entity manages this mapping of virtual piece of land with this physical earth. Moreover, what will be the legal implication of owing this digital land without being mapped onto real earth and how the market value will be determined?

### Representation of Same Digital Asset Over Various NFT Platforms

Another issue is that the same piece of virtual land, for instance, Eiffel Tower, can be recorded as digital asset on different NFT platforms. For instance, an NFT for Eiffel Tower can be minted over NextEarth, Earth2, Decentraland, and Sandbox. One future research direction is how to transfer one NFT minted on a platform to another? How multiple NFTs of the same digital asset, Eiffel Tower in this case, can be considered in real world and what will be its implication in terms of true ownership of the digital land?

### NFTs for Small Physical Items

NFTs carry value because of their uniqueness, non-interchangeability, and scarcity. But when it comes to small physical items and its representation as NFT, this may create technical challenges in terms of scalability (Park, A., et al, 2022). Even in the case of small digital assets – such as skins, weapons, clothes - a huge volume of digital assets can be minted on blockchain platform. Keeping track of ownership record for all such small physical and digital asset will create huge number of transactions on the blockchain, thus increasing the size of the blockchain, and reducing the block addition time and/or transaction speed. The question will be how much gain it will provide if central entity is removed and NFT is minted for such small physical and digital assets?

### Scams and Frauds in the NFT market

Recent studies (Kshetri, N., 2022) shows that there are several threats associated with minting, investing, and selling NFTs without the permission of the actual owner of the asset. One future research direction is to empirically investigate the impact of scams and frauds in the NFT market and devise some solutions to mitigate such scams and frauds.

## VII. CONCLUSION

Non-Fungible Tokens are a very new and rapidly emerging technology governed by the use of Smart Contracts that binds a digital asset with Ethereum blockchain. And since its bound with a blockchain, this contract enjoys the features of immutability and quick traceability which no other technology provides. A thorough analysis of scientific literature reveals that researchers are not too confident about the legitimacy of this form of tokenization. Although they do find its application in the digital rights space, especially intellectual properties like digital art, music, and assets in virtual worlds like Metaverse and Decentraland, they are still skeptical about the continuation of NFTs as a useful tool for this purpose. Gold rush around NFTs, as apparent by astronomically high value of its trades, however, tells a different story. This hype, though, is considered, by most authors, as a bubble that is soon going to burst and this major correction will leave many in possession of highly priced NFTs with almost no inherent value. Having said that, the technology behind NFTs appears so have an approval from the researchers especially the feature of smart contract that has many other applications then just registering a digital art or music on the blockchain. Following, therefore, could be concluded:

1. NFTs provide the luxury of immutability and traceability which has found its applications in the digital rights and digital assets ownership space. This is especially helpful for digital artists who previously had no way of generating a consistent income out of their work.
2. The mechanism of 'Smart Contract', is the key here because these contracts can be used to find many new applications of this technology.
3. These smart contracts protected digital rights make NFTs very useful for trading in virtual worlds like Metaverse and Decentraland.
4. NFTs, however, must be adopted with caution as it still has to test its long-term resilience and the current hype around its value is bound to go through a major correction.

As a concluding remark, and answer to the questions that this study posed in the beginning, NFTs are here to stay although not necessarily in the form of highly priced digital art only. Applications of NFTs could be innumerable and in any field where an immutable and traceable identification and ownership of assets is needed. A few areas in which further research on NFTs in business and management may spread could be product traceability in supply chain management, intellectual property rights protection, and online trades in virtual worlds. This last application could have immense potential for NFTs to thrive and give ways of innumerable fields of research. Imagine a virtual world existing, with its own economy, parallel to the real material world. Whatever ownership and trade related areas of study we see in this real world will also exist in one form or the other in that virtual world thus opening up a vast avenue of totally novel ideas, concepts, applications, and of course, research.